\documentclass[a4paper,onecolumn,11pt,accepted=2026-11-23]{quantumarticle}
\pdfoutput=1
\usepackage[utf8]{inputenc}
\usepackage[english]{babel}
\usepackage[T1]{fontenc}
\usepackage{amsmath}

\usepackage{tikz}

\usepackage[numbers,sort&compress]{natbib}
\usepackage[colorlinks=true,linkcolor=blue,citecolor=blue,urlcolor=blue]{hyperref}
\usepackage{multirow}

\usepackage{graphicx}
\usepackage{amsmath,amssymb}
\usepackage{dsfont}
\usepackage{xcolor}
\usepackage{amsthm}
\usepackage{float}
\usepackage{braket}

\theoremstyle{remark}

\makeatletter
\def\l@subsection#1#2{}
\def\l@subsubsection#1#2{}
\makeatother

\begin{document}

\title{Dynamical quantum codes and logic gates on a lattice with sparse connectivity}
\author{Dominic J. Williamson}
\thanks{Current address: School of Physics, The University of Sydney, NSW 2006, Australia}
\affiliation{IBM Quantum, IBM Almaden Research Center, San Jose, CA 95120, USA}
\author{Bence Het\'{e}nyi}
\affiliation{IBM Quantum, T. J. Watson Research Center, Yorktown Heights, New York 10598, USA}
\affiliation{IBM Quantum, IBM Research Zurich, 8803 Ruschlikon, Switzerland}
\date{}

\begin{abstract}
\noindent
We introduce several dynamical schemes that take advantage of mid-circuit measurement and nearest-neighbor gates on a lattice with maximum vertex degree three to implement topological codes and perform logic gates between them. 
We first review examples of Floquet codes and their implementation with nearest-neighbor gates and ancillary qubits. 
Next, we describe implementations of these Floquet codes that make use of the ancillary qubits to reset all qubits every measurement cycle.
We then show how switching the role of data and ancilla qubits allows a pair of Floquet codes to be implemented simultaneously. 
We describe how to perform a logical Clifford gate to entangle a pair of Floquet codes that are implemented in this way. 
Finally, we show how switching between the color code and a pair of Floquet codes, via a depth-two circuit followed by mid-circuit measurement, can be used to perform syndrome extraction for the color code. 
\end{abstract}

\maketitle

{\footnotesize
\tableofcontents
}

\section{Introduction} 
\label{sec:Introduction}

Quantum error correction is a crucial requirement for scaling up fault-tolerant quantum computers. 
Planar architectures offer the advantage of simplicity and locality, at the cost of easy implementation of high-rate codes~\cite{bravyi2010tradeoffs,bravyi2011subsystem}. 
Recent work on dynamical codes, including Floquet codes~\cite{Hastings2021,Paetznick2022,Vuillot2021,Gidney2021Honeycomb,Haah2022,Aasen2022,Ellison2023,Aasen2023,Davydova2023,Kesselring2024,GransSamuelsson2024,Davydova2024QAC,Claes2025,Setiawan2025}, has further eased the requirements of implementing topological codes on architectures with nearest-neighbor connectivity in two spatial dimensions~\cite{McEwen2023,hesner2024using,Benito2025,harper2025characterising}.

Error correction protocols on low-connectivity superconducting qubit devices are typically limited to small-scale experiments~\cite{harper2019fault,gupta2024encoding} or codes without a threshold~\cite{sundaresan2023demonstrating,Hetenyi2024}.
In this work, we introduce dynamical protocols for topological quantum error correction that use qubits laid out on the vertices and edges of a hexagonal honeycomb (HH) lattice with nearest-neighbor gates and single-qubit mid-circuit measurements with global classical communication and feedforward operations. We provide numerical evidence that these protocols exhibit a finite threshold under circuit-level noise, while naturally allowing for a periodic leakage reset~\cite{mcewen2021removing}.

Here, we introduce dynamical protocols to implement topological codes that make full use of the qubits and gates available on the vertices and edges of the HH lattice. 
By using both vertex and edge qubits to support quantum codes, we find dynamical protocols that can encode more qubits than standard implementations of the surface code and its Floquet code variants by interleaving multiple codes on the same lattice. 
We introduce low-depth adaptive circuits for switching between our dynamical codes and the color code, allowing one to implement the color code without requiring square-grid connectivity~\cite{gidney2023new,lacroix2025scaling}. 
We show how to perform transversal logic gates between the interleaved dynamical codes via low-depth circuits of native single- and two-qubit gates.

The manuscript is organized as follows. 
In section \ref{sec:2} we review Floquet codes and their implementation on the HH lattice. 
In section \ref{sec:3} we introduce a new dancing implementation of the Floquet codes in which all qubits are reset each measurement cycle. 
In section \ref{sec:5} we demonstrate that the dancing implementation of the Floquet codes allows a pair of Floquet codes to be implemented simultaneously in a duet, which we call the double Floquet code, on which an entangling logical Clifford gate can be performed. 
In section \ref{sec:6} we introduce an adaptive circuit to switch between the Double Floquet code and the color code and show that this can be used to realize a dynamical implementation of the color code allowing for its transversal logical Clifford operations to be performed.
In the appendix we comment briefly on an additional implementation of the surface code on the HH lattice.

\section{Floquet codes} 
\label{sec:2}

Floquet codes implement a topological quantum memory via a repetitive sequence of two-body nearest-neighbor measurements on a two-dimensional lattice of qubits~\cite{Hastings2021}. By constantly switching between instantaneous stabilizer codes, they provide a more flexible framework for quantum error correction within the stabilizer formalism than static stabilizer codes. 

Below, we depict a non-destructive projective $XX$ measurement on a pair of qubits. 
\begin{align}
\vcenter{\hbox{\includegraphics[page=3]{Figures}}} 
\end{align}
In our circuit diagrams, time runs up the page.
We assume the classical measurement outcome is recorded and stored. 
We do not explicitly depict the path of the classical data produced by the measurement outcome in our simplified notation. 
In all sections of this manuscript, there is no need to perform quantum feed forward correction operations as deferred corrections can be simulated efficiently within the stabilizer formalism and applied at a later time or absorbed into a change of basis on the final measurement step~\cite{Gottesman1997}. 

We consider implementing dynamical codes, including Floquet codes, on the HH lattice. 
Below we depict a patch of the HH lattice. 
\begin{align}
\vcenter{\hbox{\includegraphics[page=1]{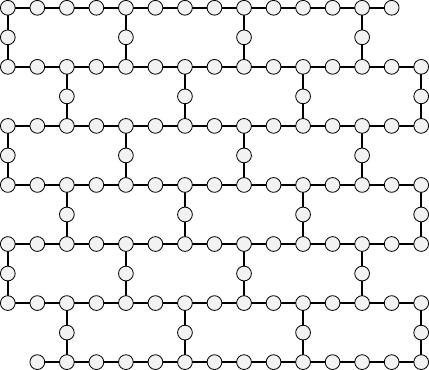}}} 
\end{align}
While this physical layout appears to be quite restrictive, we make use of dynamical rather than static codes to implement topological quantum error correction and logical operations. 

On the HH lattice, two body measurements on nearest-neighbor vertex qubits are mediated by an ancilla qubit on the edge connecting the vertices. 
The standard circuit to measure $XX$ is depicted below.
\begin{align}
\vcenter{\hbox{\includegraphics[page=4]{Figures}}} 
\end{align}
The generalization to $YY$ and $ZZ$ is straightforward by conjugating the vertex qubits with $XS$ or $H$ gates respectively. 
Where $XS$ implements a $\pi$ rotation of the Bloch sphere about the line $x+y$ and $H$ is the Hadamard gate. 

To describe the Floquet code measurement schedule it is convenient to introduce a plaquette 3-coloring of the HH lattice. 
This induces a 3-coloring of the edges as shown below. 
\begin{align}
\vcenter{\hbox{\includegraphics[page=2]{Figures}}}
\label{fig:Floquet_lattice}
\end{align}
The measurement sequence of the Hastings-Haah Floquet code (not taking into account boundary conditions~\cite{Haah2022}) consists of repeating the fundamental length three measurement cycle shown below~\cite{Hastings2021}.
\begin{align}
\vcenter{\hbox{\includegraphics[page=8]{Figures}}} 
\end{align}
Here the color appearing at each step indicates that all HH edges with the same color are measured at that step.
The Pauli label in the box below each step indicates the operator that is measured on the vertices adjacent to each appropriately colored edge. 
At a given time step, we refer to the edges that are measured as the \textit{active edges} at that time step. 

For the remainder of this work, we focus on the CSS Floquet code~\cite{Davydova2023,Kesselring2024}. 
Its measurement sequence consists of repeating the length-six cycle shown below,
\begin{align}
\label{eq:CSSFC}
\vcenter{\hbox{\includegraphics[page=9]{Figures}}} 
\end{align}
where the notation is similar to the previous equation.

To incorporate open boundary conditions we introduce further decoration on the boundary of the lattice. 
First, we include additional appropriately colored edges to ensure all boundary vertices are trivalent. 
We remark that these additional boundary edges are introduced for convenience and do not need to support physical qubits for the protocols described in this section. 
Next, we extend the coloring onto all truncated plaquettes. 
Finally we remove four corner plaquettes to form disjoint groups of horizontal and vertical truncated boundary plaquettes. 
We label the horizontal truncated boundary plaquettes by $Z$ and the vertical truncated boundary plaquettes by $X$. 
There are at least three reasonable choices for removing the corner plaquettes shown below
\begin{align}
\vcenter{\hbox{\includegraphics[page=26]{Figures}}}
\label{fig:HHX_dangling}
\end{align}
\begin{align}
\vcenter{\hbox{\includegraphics[page=33]{Figures}}} 
\end{align}
\begin{align}
\vcenter{\hbox{\includegraphics[page=34]{Figures}}} 
\end{align}
For the remainder of this section we use the first choice depicted above. 
In the presence of a boundary, the measurement steps that define the Floquet code are modified. 
For example, during a red edge $XX$ check step we additionally measure $X$ on each vertex qubit that is connected to a dangling red edge on an $X$-type (vertical) boundary. 
We remark that vertex qubits connected to dangling red edges on $Z$-type (horizontal) boundaries are not measured during this step. 
The above rule generalizes in a straightforward way to the other colors and Pauli $ZZ$ measurements, see Ref.~\cite{Kesselring2024}. 
With this modification to each measurement step, the sequence keeps the same form as above in Eq.~\eqref{eq:CSSFC}. 
This sequence of measurements can be viewed as switching between instances of the surface code on \textit{effective edge qubits} on the active edges at each time step~\cite{Hastings2021}. 
The effective edge qubits at each step are defined by two dimensional subspaces, such as $XX=\pm 1$, on pairs of vertex qubits that are projected onto by edge measurements, such as $XX$ on red edges, performed at that step.
The instantaneous surface codes that occur at each step of the measurement sequence in Eq.~\eqref{eq:CSSFC} are depicted below. 
\begin{align*}
&\vcenter{\hbox{\includegraphics[page=27]{Figures}}} 
&
\vcenter{\hbox{\includegraphics[page=28]{Figures}}} 
\end{align*}
\begin{align*}
&\vcenter{\hbox{\includegraphics[page=29]{Figures}}} 
&
\vcenter{\hbox{\includegraphics[page=30]{Figures}}} 
\end{align*}
\begin{align*}
&\vcenter{\hbox{\includegraphics[page=31]{Figures}}} 
&
\vcenter{\hbox{\includegraphics[page=32]{Figures}}} 
\end{align*}
The basis for the surface codes depicted above, on the triangular superlattices formed by the effective edge qubits, alternate according to the basis of the measurements that define the effective edge qubits. 
The first, third, and fifth steps appearing in the left-hand column have $X$-type vertex checks $\prod_{e\ni v} X_e$ and $Z$-type plaquette checks $\prod_{e \in p} Z_e$, where $e,$ $v,$ and $p$, refer to edges, vertices, and plaquettes, of the triangular superlattices. The second, fourth and sixth steps appearing in the right-hand column have $Z$-type vertex checks $\prod_{e\ni v} Z_e$ and $X$-type plaquette checks $\prod_{e \in p} X_e$, which are flipped compared to the standard formulation~\cite{qdouble,Bravyi1998}.
From this we see that the instantaneous surface codes have an effective minimum distance of three. 
We remark that the physical fault distance can be larger than this effective distance as each effective edge qubit is made up of a pair of physical qubits.

\begin{figure}
    \centering
    \includegraphics[width=0.5\linewidth]{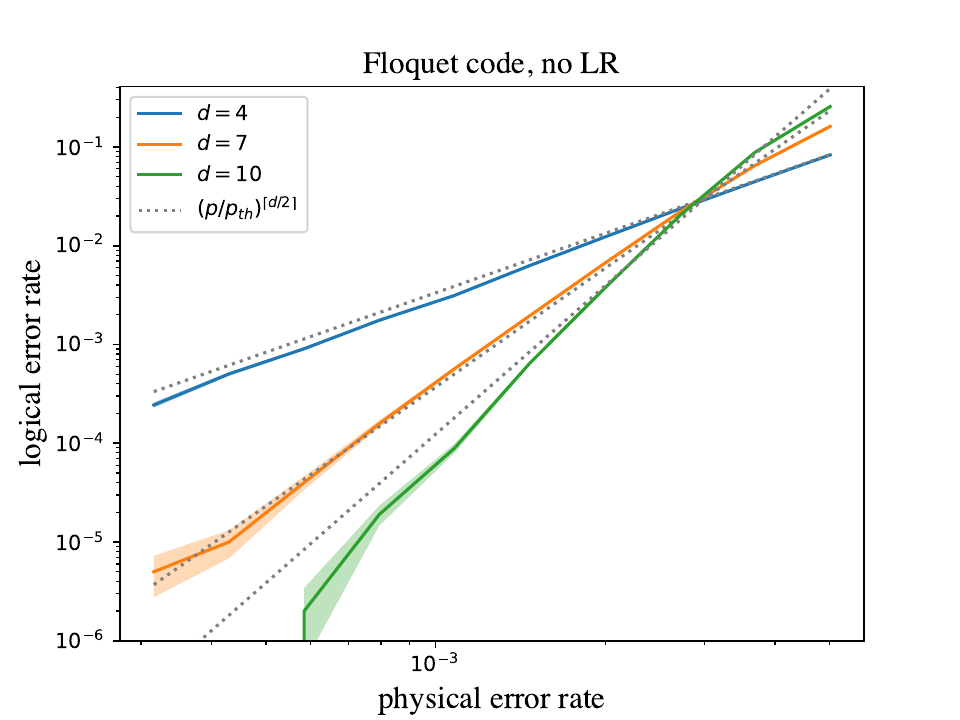}
    \caption{Threshold plot of the Floquet code under circuit-level noise. The logical $\ket{0}$ state is prepared followed by $T=1,2,3$ full (6-round) cycles of stabilizer measurements for $d=4,7,10$, respectively.}
    \label{fig:Floquet_threshold}
\end{figure}

Fig.~\ref{fig:Floquet_threshold} shows the logical error rate of the Floquet code as a function of the physical error rate, assuming circuit-level noise (including data qubit depolarizing noise during ancilla measurements) and using PyMatching~\cite{Higgott2022} for the decoding. 
The sampling of the measurements and the construction of the decoding graphs in all numerical simulations reported in this work was done using stim~\cite{Gidney2021}. 
The spacetime detectors in the decoding graph are equivalent to those in Ref.~\cite{Kesselring2024}. 
The threshold is $p_\text{th}\sim 0.3\%$ similar to the Floquet code introduced in Ref.~\cite{Kesselring2024}, even though the boundary conditions are slightly different. 
To increase the code distance in the numerical simulations, we added three columns and two rows to the HH lattice to maintain the same colors at the boundary. Hence, the code distances alternate between even and odd in the figure.

\section{Dancing Floquet code} 
\label{sec:3}

Despite its dynamical nature, the Floquet code implementation in the previous section has a static set of data qubits. 
This makes it susceptible to leakage from the physical qubit subspace over a sufficiently long time period. 
In this section, we present an alternate implementation of the Floquet code that makes use of the edge qubits to swap the role of data and ancilla qubits during every measurement step. 
This ensures that all qubits are reset during a Floquet code cycle. 

The basic idea of the dancing Floquet code is to move the effective qubit subspace on each active edge during a measurement step onto the physical edge qubits that served as ancillas in the previous section. 
To accomplish this, we introduce an alternative measurement of the $XX$ operator below.
\begin{align}
\label{eq:MeasureIn}
\vcenter{\hbox{\includegraphics[page=5]{Figures}}} 
\end{align}
In the $XII=+1,IIX=+1,$ subspace this maps $\ket{++} \mapsto \ket{+}$ and $\ket{--} \mapsto \ket{-}$. 
The effect of a $(-1)$ measurement result of $XII$ or $IIX$ switches the corresponding input state from $\ket{+}$ to $\ket{-}$ in the above mapping.

To reinitialize the vertex qubits in the $XX=+1$ subspace we use the circuit below.
\begin{align}
\label{eq:ResetOut}
\vcenter{\hbox{\includegraphics[page=6]{Figures}}} 
\end{align}
For the $Z=+1$ outcome this maps $\ket{+} \mapsto \ket{++}$ and $\ket{-} \mapsto \ket{--}$. 
A $Z=-1$ outcome can be effectively corrected to a $Z=+1$ outcome by applying $X$ to either of the outgoing qubits. 
If one or both of the first and third qubit are initialized in $\ket{-}$ rather than $\ket{+}$ this circuit undoes the mapping for corresponding $(-1)$ measurement results of $XII$ or $IIX$ in Eq.~\eqref{eq:MeasureIn}. 

Combining the above circuits we can measure $XX$ and reinitialize the vertex qubits by switching the data onto the edge qubit. 
We call this a leakage-reducing measurement of $XX$. 
\begin{align}
\vcenter{\hbox{\includegraphics[page=7]{Figures}}} 
\label{eq:LeakageReducingMsmnt}
\end{align}
Here, for each $X$ measurement we choose to reinitialize the measured qubit in the $\ket{+}$ state if $X=+1$ is observed, and the $\ket{-}$ state if $X=-1$ is observed. 
We further apply an $X^\sigma$ operator to the first qubit when a $Z=(-1)^\sigma$ outcome is observed. 

Similar circuits can be found to measure $YY$ and $ZZ$ (or any pair of Pauli matrices) by changing basis on the first and third qubtis above. 
For $ZZ$ measurements, we find it convenient to change the basis of all three qubits via a Hadamard operator. 
We use this convention throughout the manuscript.

\begin{figure}
    \centering
    \includegraphics[width=0.9\linewidth]{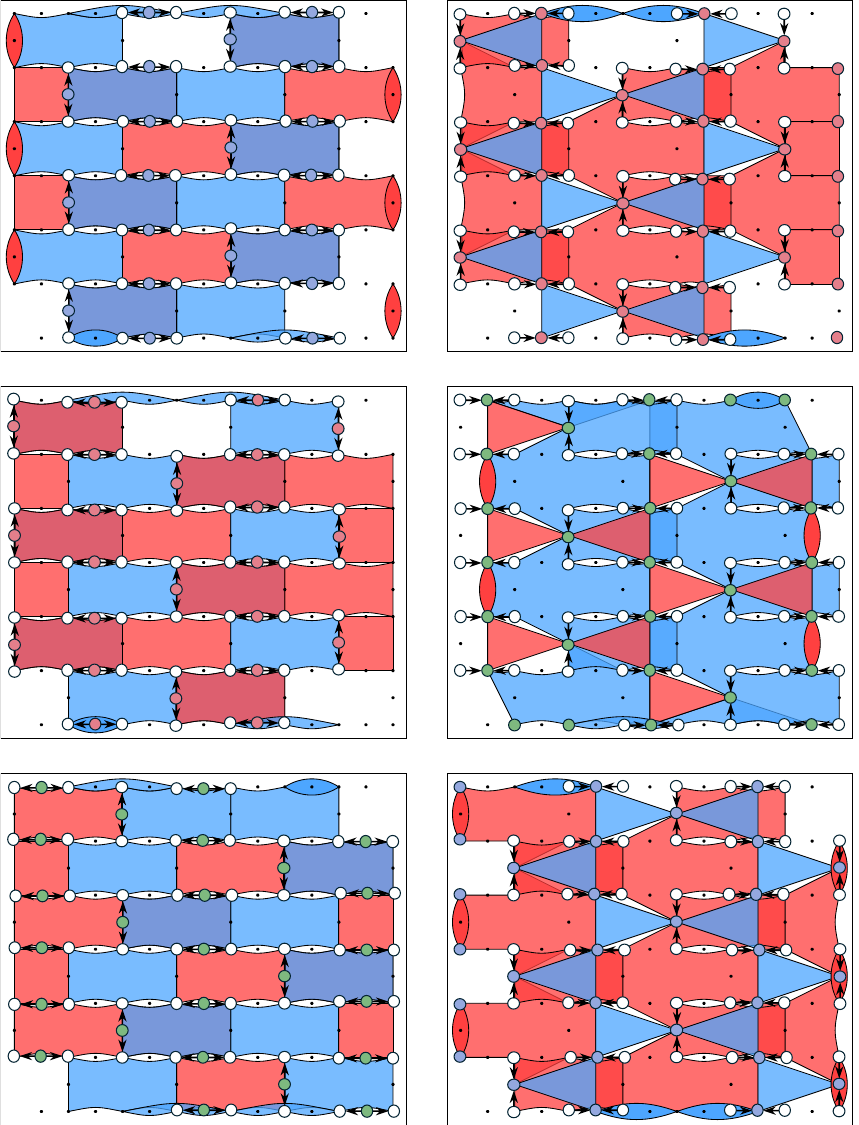}
    \caption{Detector slice diagrams of the Floquet code with leakage-reducing parity measurements along every link. Red (blue) regions correspond to X (Z) regions. The first diagram represents the detectors after the expansion (Eq.\eqref{eq:ResetOut}) of blue ZZ links. Then the red XX, and green ZZ measurements follow, and finally the contraction (Eq.\eqref{eq:MeasureIn}) of the blue XX. The second half of the cycle is not shown. Snapshots on the right column correspond to the later three steps in Eqs.~\eqref{eq:CSSFC} which were depicted in the section above (here the left and right boundaries include additional complete links that are part of the lattice). Single nodes represent single-qubit measurements on incomplete links.
}
    \label{fig:FloquetLR_det_slices}
\end{figure}

The measurement procedure above divides the measurement of each set of edge checks into two steps. 
In the first step the information is moved onto the active edge qubits and the vertex qubits are measured. 
This is depicted on the far left in Eq.~\eqref{eq:WanderingFloquetSequence} below, via a colored edge with arrows pointing in. 
Next, the information is moved back onto the vertex qubits and the active edge qubits are measured out. 
This is depicted following the second arrow in Eq.~\eqref{eq:WanderingFloquetSequence} below, via the colored edge with arrows pointing out. 
Below we have depicted a general Floquet measurement sequence that can form either the Hastings-Haah Floquet code or half of the CSS Floquet code cycle depending on the choice of the Pauli matrices $P_i$. 
\begin{align}
\vcenter{\hbox{\includegraphics[page=10]{Figures}}} 
\label{eq:WanderingFloquetSequence}
\end{align}
We remark that in the presence of boundaries, the boundary qubits are already being reset once per Floquet cycle without further modification. Fig.~\ref{fig:FloquetLR_det_slices} shows the deformation of the Floquet code stabilizers to the intermediate surface code state where the data qubits live on the edges.

\begin{figure}
    \centering
    \includegraphics[width=0.45\linewidth]{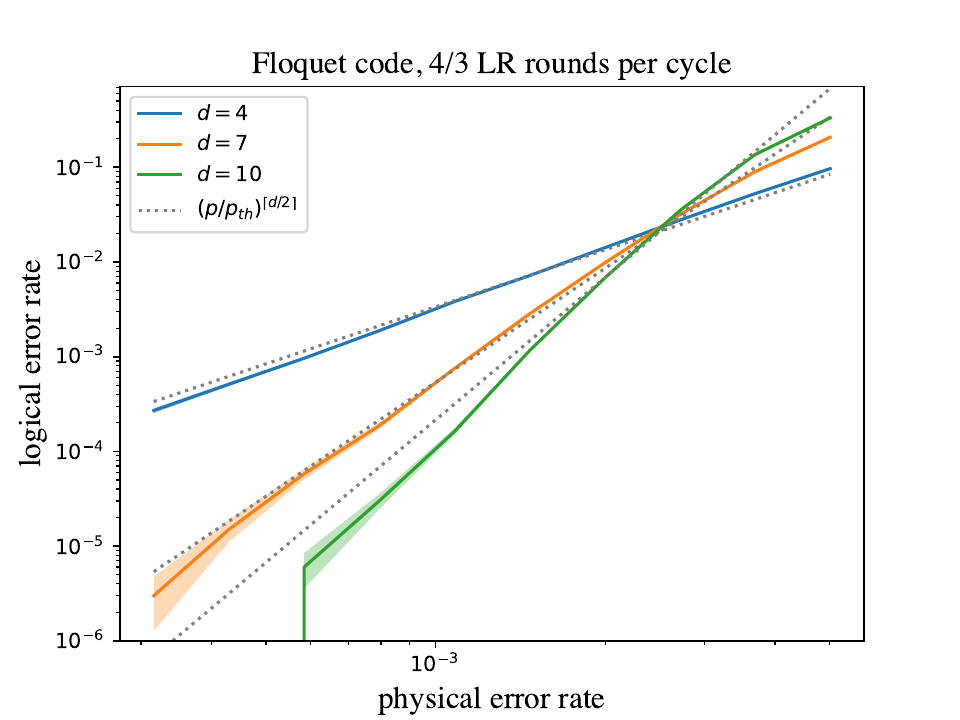}
    \includegraphics[width=0.45\linewidth]{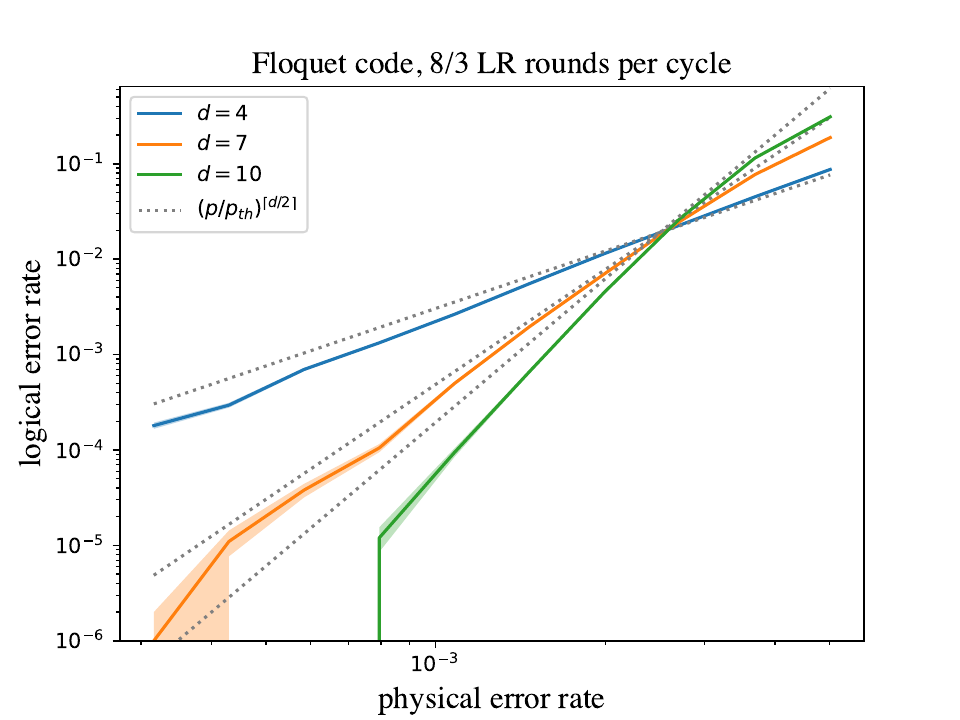}
    \includegraphics[width=0.45\linewidth]{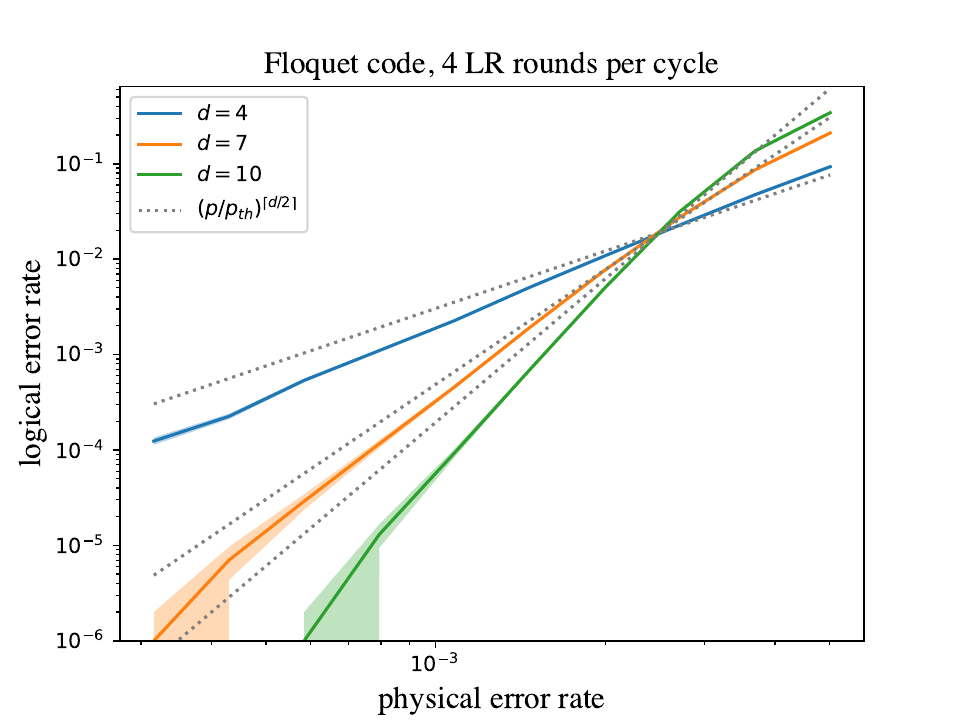}
    \includegraphics[width=0.45\linewidth]{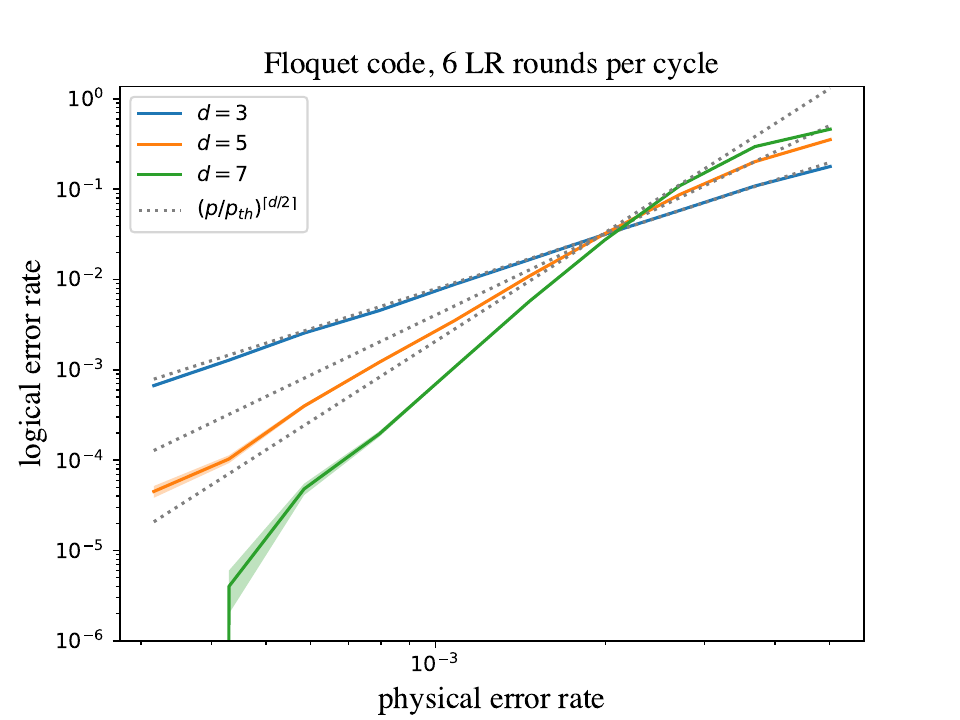}
    \caption{Threshold plots of the dancing Floquet code under circuit-level noise. Top left: distance preserving version resetting every qubit at least once (on average 4/3 times) per 6-round cycle; Top right: resetting data qubits at least twice twice (on average 8/3 times) per cycle; Bottom left: resetting data qubits four times per cycle; Bottom right: distance-reducing version resetting every data qubit in every round. The colors correspond to the same physical system size (as in Fig.~\ref{fig:Floquet_threshold}). The logical $\ket{0}$ state is prepared followed by $T=1,2,3$ full cycles of stabilizer measurements for increasing system size.}
    \label{fig:Floquet_LR_threshold}
\end{figure}

Above we have described applying leakage-reducing parity measurements in place of every standard edge measurement in a Floquet code. 
It is also possible to only replace a subset of the standard edge measurements in a Floquet code with leakage-reducing measurements. 
It is desirable to only deploy leakage-reducing parity measurements on edges that are not aligned with a logical operator of the same Pauli type as the edge measurements. 
This is because leakage-reducing parity measurements of $XX$ on edges that align with an $X$-type logical operator can lead to a reduction in the code distance, and similarly for $Y,Z$. 
By deploying standard measurements on at least one third of the links in every round, we were able to preserve the code distance while resetting every data qubit four times in every six rounds (see Fig.~\ref{fig:Floquet_LR_threshold} bottom right for the corresponding threshold plot). 
Furthermore, we were able to perform every edge measurement while preserving the code distance in up to four rounds, while still resetting every data qubit at least once over a Floquet cycle (see Fig.~\ref{fig:Floquet_LR_threshold} top left). 
Interestingly, while deploying leakage reduction appears to reduce the threshold ($p_\text{th} \sim 0.25\%$) compared to the standard Floquet code, increasing the number of data qubit resets per Floquet cycle appears to preserve this threshold and even slightly improve the sub-threshold scaling. This might be attributed to the number of rounds used in the simulations, but we did not investigate this aspect in detail. The spacetime detectors used to find the numerical results in Fig.~\ref{fig:Floquet_LR_threshold} are similar to those from Ref.~\cite{Kesselring2024} but are modified to incorporate the additional measurements and qubit resets introduced by the leakage-reducing parity measurements. 

Finally, we compare the distance-preserving dancing Floquet codes with the distance-reduced version, where every link measurement is performed using the leakage-reducing parity measurement (see Fig.~\ref{fig:Floquet_LR_threshold} bottom right). The three cases ($d=3,5,7$) correspond to the same lattice size as for the other plots, but with a reduced code distance. Interestingly, the distance-reduced version has a more favorable subthreshold scaling than is expected from the code distance, in the regime of practically relevant error rates ($p\sim10^{-3}$). Therefore, the distance-reduced version may not represent as much of a compromise to performance as one would naively expect.

\subsection*{Dancing Floquet code with qubit switching} 
\label{sec:4}

Since measuring and resetting qubits may take a relatively long time, we now introduce an extra step to the above procedure that allows this to be done in a more parallel fashion. 
To achieve this we introduce an additional round of swap gates between each edge measurement and reinitiailization step. 
This is depicted schematically below. 
\begin{align}
\vcenter{\hbox{\includegraphics[page=11]{Figures}}} 
\end{align}
Here, the second step out of every three steps corresponds to swapping vertex and edge qubits to shift the location of qubits being measured from vertices to edges. The purpose of this step is to free up vertex qubits for immediate use while performing measurements on edge qubits that would otherwise be unused at that particular time step. 
The swaps indicated above are intended to be performed before the vertex-qubit measurement involved in the in-arrow steps. 
This is to ensure that the measurement is performed once the vertex qubit states have been swapped onto adjacent edge qubits. 
The swaps, between the red measurement and reinitialization steps, exchange the vertex qubits on one sublattice with adjacent green edge qubits and the vertex qubits on the other sublattice with the adjacent blue edge qubits. 
The swaps between green, or blue, measurement and reinitialization steps proceed similarly. 
The swaps bring edge qubits in to serve as vertex qubits for the following out-arrow edge, without the need to wait until the measurement is completed. 
This requires a slight modification to the measurement procedure in Eq.~\eqref{eq:LeakageReducingMsmnt} as the measurement outcomes may not yet be known. 
Instead one can always reinitialize the vertex qubits in the $\ket{+}$ state and later apply a $Z$ operator to the respective qubits where an $X=-1$ outcome is observed, and similarly for measurements of $Y,Z$. 

\section{Double Floquet code} 
\label{sec:5}

In this section we show that physically moving the quantum state onto active edge qubits in the Floquet code has the advantage that it allows a pair of Floquet codes to be run in tandem on the same HH lattice. 
The basic idea is that while the first Floquet code lives on the red edge qubits (for example) the second Floquet code advances from the green edge qubits to the vertices and then onto the blue edge qubits. 
The first Floquet code can then advance from the red edges to the vertices and then to the green edges, and so on. 
The interlaced measurement sequence is depicted below, where the steps of the first Floquet code are on top and the second Floquet code on the bottom. 
\begin{align}
\vcenter{\hbox{\includegraphics[page=12]{Figures}}} 
\label{eq:DoubledFloquetSequence}
\end{align}
In the presence of open boundaries the above sequence requires additional physical edge qubits where unmeasured boundary qubits can be moved to avoid collisions between the Floquet codes. 
The simplest solution is to include the additional edge qubits depicted in Eq.~\eqref{fig:HHX_dangling}. 
However, it is not strictly necessary to include all such qubits. 
For example, for the Floquet sequence in Eq.~\eqref{eq:DoubledFloquetSequence}, the blue boundary on the left in Eq.~\eqref{fig:HHX_dangling} does not require additional dangling physical qubits. 
This is because additional swaps can be used to advance the second Floquet code from red edges to green edges while the first Floquet code is supported on blue edges and boundary vertices without additional boundary blue edges. 
Analogous statements hold for other appropriate choices of Floquet cycles, boundary colors, and orientations. 

\begin{figure}[h]
    \centering
    \includegraphics[width=0.5\linewidth]{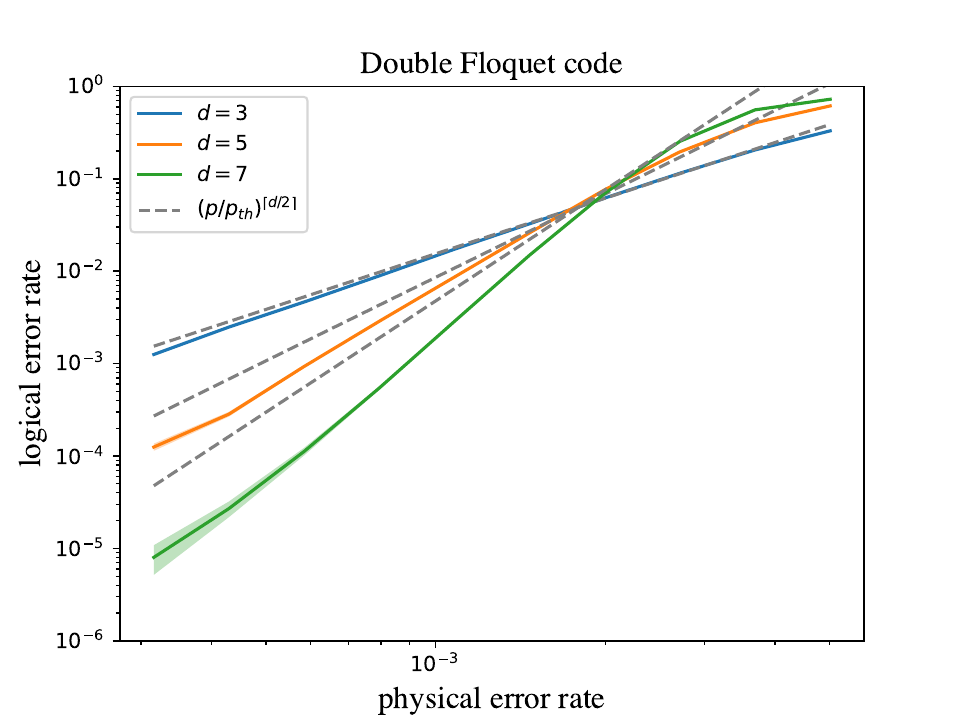}
    \caption{Threshold plot of the Double Floquet code under circuit-level noise. The logical $\ket{00}$ state is prepared followed by $T=1,2,3$ full cycle of stabilizer measurements for $d=3,5,7$, respectively.}
    \label{fig:double_Floquet_threshold}
\end{figure}

Simulating the Double Floquet code under circuit-level noise (see Fig.~\ref{fig:double_Floquet_threshold}), we have found similar favourable sub-threshold scaling as for the non-distance-preserving version of the dancing Floquet code in Fig.~\ref{fig:Floquet_LR_threshold}~(bottom-right). 
On the other hand, the value of the threshold ($p_\text{th}\sim 0.2\%$) is reduced due to the doubled circuit depth and gate count per Floquet cycle.
In our simulations, the dangling edges shown in Eq.~\eqref{fig:HHX_dangling} are included as physical qubits to avoid collisions between the Floquet codes. 
The spacetime detectors of the Double Floquet code are similar to those of the dancing Floquet code in the previous section. 

\subsection*{Transversal gates on the Double Floquet Code} 

The Double Floquet code facilitates the application of logical Clifford gates such as CNOT. 
For example, consider the surface codes that appear at the red $XX$ and blue $XX$ Floquet code measurement steps. 
After measuring out the edge qubits on the bottom most row of the red surface code, and top most row of the blue surface code, in the $Z$ basis we are left with an identical pair of surface codes that have been shifted by 4 lattice sites relative to one another. 
\begin{align}
\vcenter{\hbox{\includegraphics[page=41]{Figures}}} 
\quad 
\vcenter{\hbox{\includegraphics[page=42]{Figures}}} 
\end{align}
Appropriate pairs of red and blue edge qubits can then be brought next to one another via a depth-three nearest-neighbor swap circuit. 
The order of the bonds on which swap gates are applied is shown below.  
First a swap is applied to the pair of qubits adjacent to each edge that is assigned a 1 label, followed by a swap on the pair of qubits adjacent to each edge assigned a 2 label, finally a swap on the pair of qubits adjacent to each edge assigned a 3 label. 
\begin{align}
\vcenter{\hbox{\includegraphics[page=43]{Figures}}} 
\end{align}
Next, a nearest neighbor transversal CNOT can be applied to implement a logical CNOT between the surface code patches to entangle their logical qubits. 
The surface code patches can then be swapped back to their original positions and the Floquet error correction cycle can continue. 
We remark that decoding this gate requires techniques beyond standard minimum-weight perfect matching~\cite{Sahay2024,Cain2024,Cain2025,Turner2025,SerraPeralta2025}.

Access to a pair of surface codes located on the same lattice patch provides a significant advantage in that it supports a bilayer architecture.
This allows for the application of transversal logic gates and it allows more flexible routing of logical qubits via lattice surgery, or swapping the physical qubits in patches of surface code. 
In particular this can be used to implement the low overhead fault-tolerance scheme proposed in Ref.~\cite{Pattison2023}. 

\section{Switching color code} 
\label{sec:6}

In this section we consider implementing the color code on the HH lattice via adaptive circuits that switch between the color code and copies of the surface code. 
This approach also allows one to switch from the Double Floquet code, introduced above, to the color code. 

The color code is defined by the simultaneous $(+1)$-eigenspace of plaquette checks, $\prod_{v \in p} X_v,$ $\prod_{v \in p} Z_v,$ supported on qubits on the vertices of a trivalent lattice with 3-colorable plaquettes~\cite{Bombin2006}. 
The algebra of the plaquette checks implies that the color code is in the $(+1)$-eigenspace of $\prod_{v \in p} Y_v$ operators, for plaquettes with a number of vertices that is divisible by four, and the $(-1)$-eigenspace of $\prod_{v \in p} Y_v$ operators, for plaquettes with a number of vertices that is divisible by two but not four. 

Consider a many-qubit stabilizer check measurement in the color code such as the six body $X$ plaquette measurement shown below. 
\begin{align}
\vcenter{\hbox{\includegraphics[page=13]{Figures}}} 
\end{align}
This measurement can be made on the HH plaquette after a depth-two circuit following a similar approach to the $XX$ measurement in Eq.~\eqref{eq:MeasureIn}. 
We depict the circuit to measure $X^{\otimes 6}$ below, using periodic boundary conditions. 
\begin{align}
\label{eq:PlaquetteDuality}
\vcenter{\hbox{\includegraphics[page=14]{Figures}}} 
\end{align}
This circuit causes a deformation from the color code to a new code. 
If the above circuit is applied to all plaquettes of a given color, the resulting code is equivalent to two copies of the surface code with boundary conditions determined by those of the initial color code. 
We remark that the above circuit is modified in a straightforward way on the truncated boundary plaquettes, the main difference being that the circuits for truncated plaquettes do not have periodic boundary conditions. 
To restore the color code, we apply the following circuit. 
\begin{align}
\vcenter{\hbox{\includegraphics[page=15]{Figures}}} 
\end{align}
The above choice of ancilla $\ket{+}$ states results in the color code being initialized in the $+1$ eigenspace of the $X$ plaquette. 
More generally, we may choose the ancilla states to match the outcome of the associated $X$ measurements that immediately precede them. 
We remark that a simple feedforward procedure allows the outcomes of the $Z$ measurements to be flipped to all $+1$, this is a simple instance of the gauging measurement procedure from Ref.~\cite{Williamson2024}. 

The above measurement and reinitialization procedure results in the following nearest-neighbor adaptive circuit that implements the measurement and reinitialization of the $X$-type plaquette operator.
\begin{align}
\vcenter{\hbox{\includegraphics[page=16]{Figures}}} 
\label{eq:CCPlaquetteMsmnt}
\end{align}
Here, following an $X=+1$ measurement result, we reinitialize the qubit in the $\ket{+}$ state, and similarly for an $X=-1$ measurement result followed by reinitializing a $\ket{-}$ state. 
We remark that the above measurement circuit generalizes simply to measuring the $Z$-type stabilizers of the color code via changing basis on all qubits in Eq.~\eqref{eq:CCPlaquetteMsmnt} using Hadamard gates. 

The measurement we have introduced above leads to a dynamical implementation of the color code on the HH lattice following a sequence of measurements shown below.
\begin{align}
\vcenter{\hbox{\includegraphics[page=17]{Figures}}} 
\end{align}
Where each plaquette color denotes the measurements of the $X$ stabilizers on all plaquettes of that color followed by the measurement of the $Z$ stabilizers on all plaquettes of that color. 
We have used STIM to verify that this circuit leads to a dynamical code with fault distance that increases with system size, although it suffers a drop form the distance $d$ of the static color code to $\geq \frac{d}{2}$. 
For this we used a sequence of triangular patch boundary conditions, see below. 
The detectors (as well as the logical observable) in this dynamical code are similar to those in a standard implementation of the color code but they involve additional measurements to account for the Pauli feed-forwards required to restore the codespace of the color code after the syndrome measurement circuit in Eq.\eqref{eq:CCPlaquetteMsmnt} (see Fig.~\ref{fig:ColorCodeDetecotrs} for details). 
This reflects the deformation of the stabilizer checks through the instantaneous steps of the Floquet code cycle depicted above.

\subsection{Triangle patch}

\begin{figure}[t]
    \centering
    \includegraphics[width=0.6\linewidth]{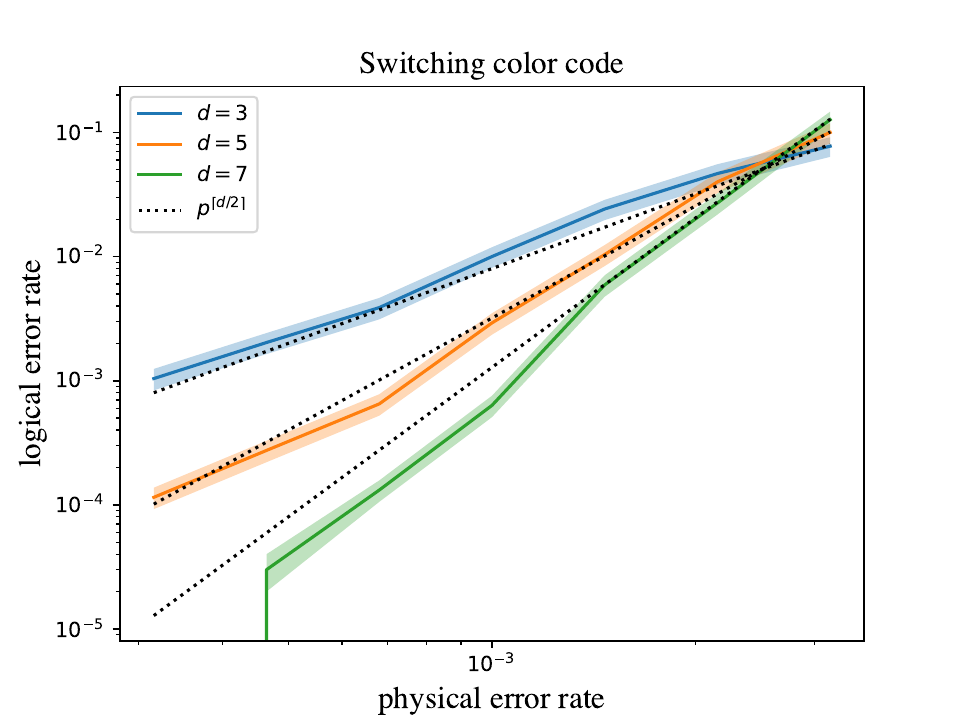}
    \caption{Threshold of the triangle-patch switching color code under circuit-level noise using a belief-propagation decision-tree decoder of Ref.~\cite{ott2025decision}. Here the logical error rate refers to the transversal version of the logical Z observable. We find a threshold value of $0.25\%$.}
    \label{fig:ColorCodethreshold}
\end{figure}

\begin{figure}[t]
    \centering
    \includegraphics[width=0.8
    \linewidth]{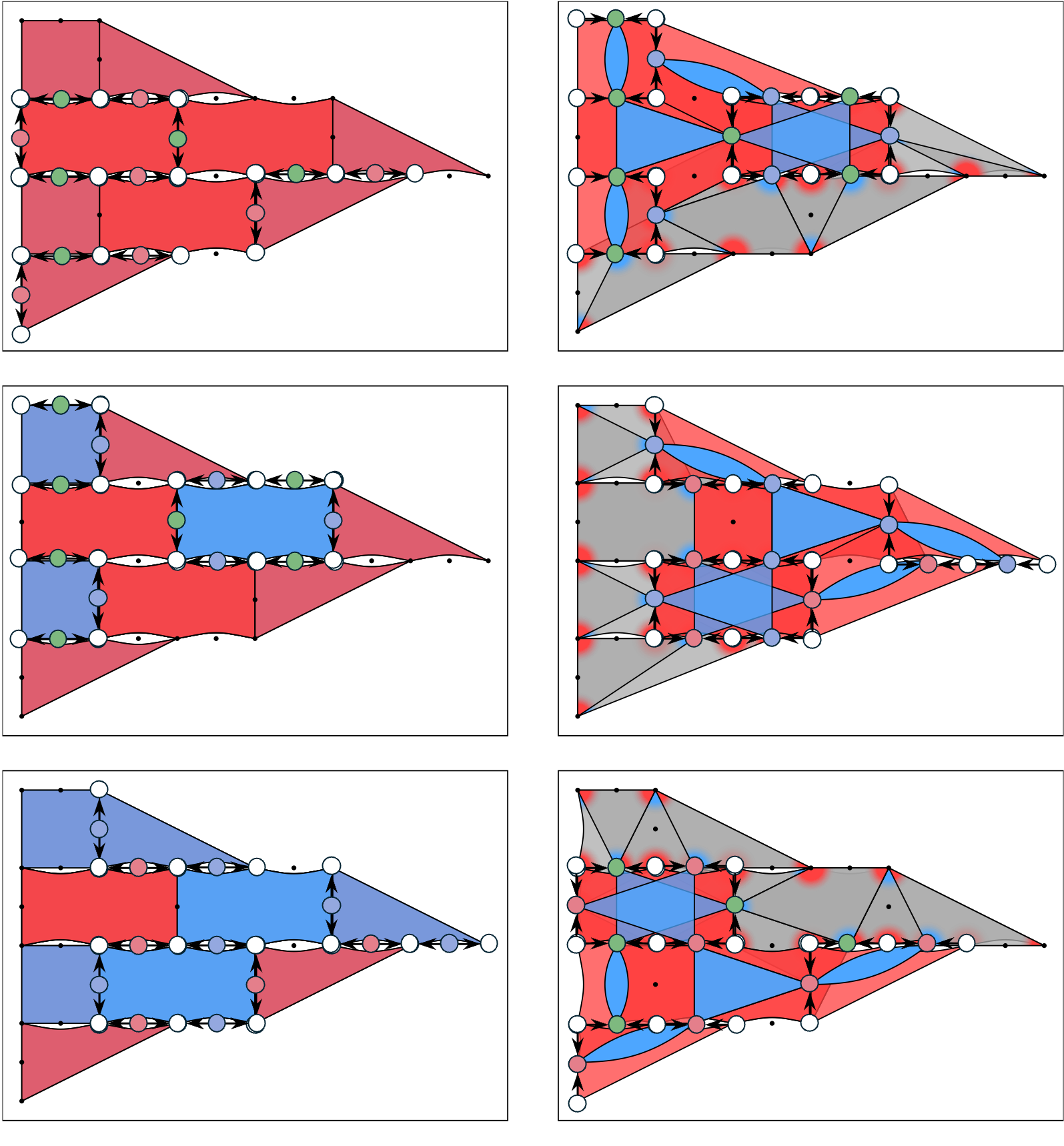}
    \caption{Detector slice diagram of the switching color code on the compact triangle patch. Red (blue) regions correspond to X (Z) regions. The first diagram shows the expansion of blue X plaquettes. Then red Z (second and third) and green Z plaquettes (fourth and fifth) are measured. Finally the contraction of the blue Z (sixth). The second half of the cycle is not shown. Grey regions appear along the fold boundary.
}
    \label{fig:ColorCodeDetecotrs}
\end{figure}

We first consider a triangular patch of the color code with one red, green, and blue, boundary. 
\begin{align}
\label{CCTriangle}
\vcenter{\hbox{\includegraphics[page=44]{Figures}}} 
\end{align}
Again, the grey qubits are not used in the protocol. 
We remark that a smaller patch of color code with the same distance fits onto the above HH lattice. 
There is also a smaller patch of color code, equivalent to Steane's code, with distance three. 
These are depicted below, and in Fig.~\ref{fig:ColorCodeDetecotrs}. 
\begin{align}
\vcenter{\hbox{\includegraphics[page=68]{Figures}}} 
\qquad
\vcenter{\hbox{\includegraphics[page=69]{Figures}}} 
\label{eq:SteaneCode}
\end{align}

The above configurations of the color code encode a single qubit and supports transversal $S$ and $H$ gates which generate the full logical Clifford group on the encoded qubit. 
This includes a logical $XS$ gate, which can be implemented by a transversal product of $XS$ gates on \textit{up} vertices, shown on the left below, and $XS^\dagger$ gates on \textit{down} vertices, shown on the right below. 
\begin{align}
\vcenter{\hbox{\includegraphics[page=57]{Figures}}} 
&&
\vcenter{\hbox{\includegraphics[page=58]{Figures}}} 
\label{eq:UpDownVertices}
\end{align}
We remark that there are three possible color configurations for the up vertices, we have depicted one choice above for simplicity, and similarly for the down vertices. 
The above combination of $XS$ and $XS^\dagger$ gates suffices to preserve the code space, as ${(XS) X (XS)^\dagger = Y}$, and ${(XS^\dagger) X (XS^\dagger)^\dagger = - Y}$. 
Hence, the transversal action maps the $X$-type plaquette operators to $Y$-type plaquette operators with the appropriate $(\pm 1)$ sign,  determined by the parity of half the number of vertices in the plaquette. 
Similarly, the precise logical action depends on the length of the logical representative up to a $(\pm 1)$ sign. 

For the remainder of this subsection we focus on the larger color code shown in Eq.~\eqref{CCTriangle} as it leads to distance three surface codes under the deformation in Eq.~\eqref{eq:PlaquetteDuality}. 
During the red measurement step the code switches into a pair of surface codes on the green and blue qubits.
Each surface code has a rough and a smooth boundary, as well as an additional fold boundary that joins the pair~\cite{kubica2015unfolding}. 
\begin{align}
\vcenter{\hbox{\includegraphics[page=45]{Figures}}} 
\end{align}
The red and green surface codes are shown separately below.
\begin{align}
\vcenter{\hbox{\includegraphics[page=48]{Figures}}} 
\quad
\vcenter{\hbox{\includegraphics[page=49]{Figures}}} 
\end{align}
Similar to the square patch protocol described below, the distance is three during the code switching. 
We remark that in this case less physical qubits are required than for the square patch, although only a single logical qubit is encoded. 
For red $X$-type plaquette measurements, the above surface codes have $X$-type vertex checks and $Z$-type plaquette checks, which can be derived by deforming the stabilizer checks of the color code through the circuit in Eq.~\eqref{eq:PlaquetteDuality}. 
Similarly during the red $Z$-type plaquette measurement step the surface codes have $Z$-type vertex checks and $X$-type plaquette checks. 
This is similar for the plaquette measurement steps below. 

During the green measurement step the code switches into a pair of surface codes on the blue and red qubits. 
Again the same assortment of boundaries appears, only their orientation has changed.
\begin{align}
\vcenter{\hbox{\includegraphics[page=46]{Figures}}} 
\end{align}
The blue and red surface codes are separated for clarity below.
\begin{align}
\vcenter{\hbox{\includegraphics[page=50]{Figures}}} 
\quad
\vcenter{\hbox{\includegraphics[page=51]{Figures}}} 
\end{align}

During the blue measurement step the code again switches into a pair of surface codes on red and green qubits with a similar configuration of boundaries. 
\begin{align}
\vcenter{\hbox{\includegraphics[page=47]{Figures}}} 
\end{align}

\begin{align}
\vcenter{\hbox{\includegraphics[page=52]{Figures}}} 
\quad
\vcenter{\hbox{\includegraphics[page=53]{Figures}}} 
\end{align}

\subsection{Square patch}

To access the transversal Clifford gates of the color code, we consider green vertical, and red or blue horizontal, boundary conditions.  
The color code with these boundary conditions encodes a pair of logical qubits and supports a transversal Hadamard gate that implements a logical Hadamard+SWAP. 
This color code also supports a logical $CZ$ gate that is implemented by transversal $XS$ $(XS^\dagger)$ gates as described in the subsection above. 

Below, on the left we have green vertical and red horizontal boundaries. 
On the right we have green vertical and blue horizontal boundaries. 
\begin{align}
\label{CCSquare}
\vcenter{\hbox{\includegraphics[page=21]{Figures}}} 
\quad \vcenter{\hbox{\includegraphics[page=25]{Figures}}} 
\end{align}
Here the grey qubits are not used in the protocol. 
We remark that we have added an extra corner qubit to both of the above color codes for convenience. 
This corner qubit and its neighbor can be moved to nearby unused grey qubits to avoid adding extra physical qubits during the protocol. 
Below we focus on the green and red boundary color code shown on the left. 

During the red $X$-type plaquette measurement step the code becomes equivalent to a pair of surface codes on the green and blue qubits. 
The green surface code has rough vertical boundaries, while the blue surface code has smooth vertical boundaries. 
The horizontal boundary is equivalent to a fold boundary, where $e$ syndromes from the green layer can pass to the blue layer, and similarly for $m$ syndromes. 
\begin{align}
\vcenter{\hbox{\includegraphics[page=23]{Figures}}} 
\end{align}
The green and blue layers are shown separately below for clarity.
\begin{align}
\vcenter{\hbox{\includegraphics[page=37]{Figures}}} 
\qquad
\vcenter{\hbox{\includegraphics[page=38]{Figures}}} 
\end{align}
These surface codes have $X$-type vertex checks and $Z$-type plaquette checks, which can be derived by deforming the stabilizer checks of the color code through the circuit in Eq.~\eqref{eq:PlaquetteDuality}. 
Similarly during the red $Z$-type plaquette measurement step the surface codes have $Z$-type vertex checks and $X$-type plaquette checks. 
This is similar for the plaquette measurement steps below. 
We remark that the above pair of surface codes has distance 3.
This is smaller than the distance of the color code due to the code deformation.

During the green plaquette measurement step the code becomes equivalent to a pair of surface codes on the blue and red qubits. 
The blue surface code has smooth horizontal boundaries, while the red surface code has rough horizontal boundaries. 
The vertical boundaries are fold boundaries during this step. 
\begin{align}
\vcenter{\hbox{\includegraphics[page=24]{Figures}}} 
\end{align}
The blue and red layers are shown separately below. 
\begin{align}
\vcenter{\hbox{\includegraphics[page=39]{Figures}}} 
\qquad
\vcenter{\hbox{\includegraphics[page=40]{Figures}}} 
\end{align}

During the blue plaquette measurement step the code is again equivalent to a pair of surface codes on the red and green qubits. 
The red surface code has rough vertical, and smooth horizontal, boundaries. 
The green surface code has rough vertical, and smooth horizontal, boundaries. 
\begin{align}
\vcenter{\hbox{\includegraphics[page=22]{Figures}}} 
\end{align}
The layers are shown separately below. 
\begin{align}
\vcenter{\hbox{\includegraphics[page=35]{Figures}}} 
\qquad
\vcenter{\hbox{\includegraphics[page=36]{Figures}}} 
\end{align}
We remark that it is possible to start from the red and green step of the Double Floquet code above and measure out qubits until the above surface codes are realized. 
One can then switch to color code to perform transversal logic gates before switching back to the surface codes. 
This results in a logical entangling $CZ$ gate on the pair of surface codes.

\section{Discussion} 
\label{sec:Discussion}

In this work, we have proposed dynamical codes that can be implemented by low-depth circuits with nearest-neighbor gates on a sparse lattice. 
By moving the data qubits between vertex and edge qubits of the HH lattice, we proposed dynamical codes that accommodate continual resetting of all qubits to suppress leakage errors.  
We further demonstrated that by making use of both the vertex and edge qubits of the HH lattice as data qubits in alternate rounds, we are able to implement multiple topological codes simultaneously on a single lattice. 
We proposed a Double Floquet code and a switching color code, both of which allow for the simple implementation of transversal logic Clifford gates. We have demonstrated thresholding behaviour for the alternative, more efficient, dynamical code implementations we propose, with comparable performance to standard implementations. 

When combined with magic state injection, the codes and gates proposed here provide a potential architecture for universal fault-tolerant quantum computing with modular HH lattice patches that can be coupled via nearest-neighbor gates along their edges. 
Gates between patches could be performed by either a sequence of swaps to move code patches around, or standard lattice surgery, combined with transversal gates within patches.

This work points to the potential gains of designing implementations of dynamical codes that are inspired by the physical connectivity and operations that are available in a specific hardware platform. 
An interesting question raised by our work is to extend our methods to the preparation of logical magic states, to complement the complete set of Clifford gates found here. A related direction is the preparation of nonabelian topological code states on the HH lattice~\cite{Tantivasadakarn2021,Tantivasadakarn2022,Iqbal2023,Ren2024,Davydova2025}. 
This raises the important question of characterizing the performance of the full planar fault-tolerant architecture derived from magic state preparation and the codes and logic gates proposed here.

\acknowledgements
We acknowledge inspiring discussions with Ben Brown, and James Wootton. The scripts and data sets used for this work are available on \href{https://github.com/hetenyib/dynamical\_codes\_on\_heavy\_hex}{GitHub}.

\section*{Author contributions}
Both authors contributed extensively to all aspects of this work.

\bibliographystyle{quantum}
\bibliography{Bibliography}

\appendix

\section{Surface code via punching out plaquettes}

Here, we briefly describe an alternative implementation of the surface code on the edges of the HH lattice. This implementation proceeds via alternating rounds of $X$-type and $Z$-type checks. The rounds of $X$-type checks consist of measuring 3-body $X$-type vertex checks on all vertices in the lattice. The rounds of $Z$-type checks cycle through three options, each option consists of measuring out edges of two colors in the $Z$-basis to infer the $Z$-type plaquette checks of the remaining color. 
A full measurement cycle consists of six rounds of checks that alternate between the $X$-type checks, and the three types of $Z$-type checks, described above. 
Implementation of the surface code memory via this scheme consists of repeating this six round syndrome extraction cycle.

\section{Distance bounds}

\begin{table}[ht]
\centering
\begin{tabular}{|c|c|c|c|}
\hline
\multicolumn{4}{|c|}{Dancing Floquet code}\\
\hline
width & basis & LR rounds per cycle & $d_\text{graph}$ \\
\hline
\multirow{10}{*}{3}
  & \multirow{5}{*}{$X$} & $0$   & \multirow{4}{*}{5} \\
  &                      & $4/3$ &                    \\
  &                      & $8/3$ &                    \\
  &                      & $4$   &                    \\
\cline{3-4}
  &                      & $6$   & 3                  \\
\cline{2-4}
  & \multirow{5}{*}{$Z$} & $0$   & \multirow{4}{*}{4} \\
  &                      & $4/3$ &                    \\
  &                      & $8/3$ &                    \\
  &                      & $4$   &                    \\
\cline{3-4}
  &                      & $6$   & 3                  \\
\hline
\multirow{10}{*}{6}
  & \multirow{5}{*}{$X$} & $0$   & \multirow{4}{*}{11} \\
  &                      & $4/3$ &                     \\
  &                      & $8/3$ &                     \\
  &                      & $4$   &                     \\
\cline{3-4}
  &                      & $6$   & 6                   \\
\cline{2-4}
  & \multirow{5}{*}{$Z$} & $0$   & \multirow{4}{*}{7} \\
  &                      & $4/3$ &                    \\
  &                      & $8/3$ &                    \\
  &                      & $4$   &                    \\
\cline{3-4}
  &                      & $6$   & 5                  \\
\hline
\multirow{10}{*}{9}
  & \multirow{5}{*}{$X$} & $0$   & \multirow{4}{*}{17} \\
  &                      & $4/3$ &                     \\
  &                      & $8/3$ &                     \\
  &                      & $4$   &                     \\
\cline{3-4}
  &                      & $6$   & 9                   \\
\cline{2-4}
  & \multirow{5}{*}{$Z$} & $0$   & \multirow{4}{*}{10} \\
  &                      & $4/3$ &                     \\
  &                      & $8/3$ &                     \\
  &                      & $4$   &                     \\
\cline{3-4}
  &                      & $6$   & 7                   \\
\hline
\end{tabular}
\caption{Graphlike distance (using stim's built-in method) of the dancing Floquet code with patch height that is twice the patch width (Fig. \eqref{fig:Floquet_lattice} corresponds to width = 3). The number of cycles used here is width/3, as for the numerical results. The noise model is identical to that of the simulations as well.}
\label{tab:distance_bounds_dancing_floquet}
\end{table}

\begin{table}[ht]
\centering
\begin{tabular}{|c|c|c|}
\hline
\multicolumn{3}{|c|}{Double Floquet code}\\
\hline
width & basis & $d_\text{graph}$ \\
\hline
\multirow{2}{*}{3} & $X$ & \multirow{2}{*}{3} \\
\cline{2-2}
                   & $Z$ &                    \\
\hline
\multirow{2}{*}{6} & $X$ & 6 \\
\cline{2-3}
                   & $Z$ & 5 \\
\hline
\multirow{2}{*}{9} & $X$ & 9 \\
\cline{2-3}
                   & $Z$ & 7 \\
\hline
\end{tabular}
\caption{Graphlike distance (using stim's built-in method) of the double Floquet code with patch height that is twice the patch width (Fig. \eqref{fig:Floquet_lattice} corresponds to width = 3). The number of cycles used here is width/3, as for the numerical results. The noise model is identical to that of the simulations as well.}
\label{tab:distance_bounds_double_floquet}
\end{table}

\begin{table}[ht]
\centering
\begin{tabular}{|c|c|c|c|}
\hline
\multicolumn{4}{|c|}{Switching color code}\\
\hline
size & basis & $d_\text{fault}$ & $d_\text{graph}$ \\
\hline
\multirow{2}{*}{5}  & $X$ & \multirow{2}{*}{3} & \multirow{2}{*}{3} \\
\cline{2-2}
                    & $Z$ &                    &                    \\
\hline
\multirow{2}{*}{9}  & $X$ & \multirow{2}{*}{5} & \multirow{2}{*}{5} \\
\cline{2-2}
                    & $Z$ &                    &                    \\
\hline
\multirow{2}{*}{13} & $X$ & \multirow{2}{*}{7} & \multirow{2}{*}{7} \\
\cline{2-2}
                    & $Z$ &                    &                    \\
\hline
\end{tabular}
\caption{Graphlike distance and the weight of the shortest undetectable logical (using stim's built-in methods) for the switching color code for different patch sizes (Fig. \eqref{eq:SteaneCode} corresponds to size = 5, i.e., the distance of the color code). The number of cycles used here is 1, as for the numerical results. The noise model is identical to that of the simulations as well. In the undetectable error search Tanner-graph nodes up to degree 6 were explored, allowing for increasing symptom degree.}
\label{tab:distance_bounds_switching_color}
\end{table}
\end{document}